\documentclass[journal]{IEEEtran}

\usepackage{amsmath,amsfonts,amssymb,mathrsfs}
\usepackage{bm}
\usepackage{microtype}
\usepackage{hyperref}

\usepackage{acro}[=v2]
\acsetup{single}

\usepackage[
style=ieee,
dashed=false,
maxbibnames=20,
minbibnames=2,
maxcitenames=1,
mincitenames=1,
backend=biber,
defernumbers=false,
sorting=none,
useprefix=false
]{biblatex}

\usepackage[dvipsnames]{xcolor}
\usepackage{subcaption}
\usepackage{graphicx}

\usepackage{algorithm}
\usepackage{algpseudocode}

\usepackage{tikz}
\usetikzlibrary{arrows,shapes,calc,decorations.pathreplacing,intersections,quotes,angles,positioning,fit,petri,through,shadows,plotmarks,spy,math}

\usepackage{pgfplots}
\usepgfplotslibrary{groupplots}

\usepackage{booktabs}
\usepackage{multirow}
\usepackage{gensymb} 

\usepackage{lipsum}

\usepackage[mathscr]{euscript}

\newcommand{\boldb}{\bm{b}}

\newcommand{\boldv}{\bm{v}}

\newcommand{\boldx}{\bm{x}}
\newcommand{\boldy}{\bm{y}}
\newcommand{\boldz}{\bm{z}}

\newcommand{\boldphi}{\bm{\phi}}

\newcommand{\boldzero}{\bm{0}}
\newcommand{\boldepsilon}{\bm{\epsilon}}
\newcommand{\boldomega}{\bm{\omega}}
\newcommand{\boldvartheta}{\bm{\vartheta}}
\newcommand{\boldtau}{\bm{\tau}}
\newcommand{\boldeta}{\bm{\eta}}

\newcommand{\boldxhat}{\hat{\boldx}}

\newcommand{\boldA}{\bm{A}}
\newcommand{\boldB}{\bm{B}}

\newcommand{\boldI}{\bm{I}}

\newcommand{\boldW}{\bm{W}}

\newcommand{\calC}{\mathcal{C}}

\newcommand{\calN}{\mathcal{N}}

\newcommand{\calX}{\mathcal{X}}
\newcommand{\calY}{\mathcal{Y}}
\newcommand{\calZ}{\mathcal{Z}}

\newcommand{\scrN}{\mathscr{N}}

\newcommand{\scrU}{\mathscr{U}}

\newcommand{\boldcalA}{\bm{\mathcal{A}}}

\newcommand{\boldcalR}{\bm{\mathcal{R}}}

\newcommand{\boldcalV}{\bm{\mathcal{V}}}
\newcommand{\boldcalW}{\bm{\mathcal{W}}}

\newcommand{\rma}{\mathrm{a}}

\newcommand{\rmc}{\mathrm{c}}

\newcommand{\rme}{\mathrm{e}}

\newcommand{\rmh}{\mathrm{h}}

\newcommand{\rmw}{\mathrm{w}}

\newcommand{\ybar}{\bar{y}}

\newcommand{\R}{\mathbb{R}}

\newcommand{\transp}{^\top}

\newcommand{\argmin}{\operatornamewithlimits{arg\,min}}

\DeclareAcronym{2D}{
	short=2-D,
	long=two-dimensional,
}

\DeclareAcronym{3D}{
	short=3-D,
	long=three-dimensional,
}

\DeclareAcronym{4D}{
	short=4-D,
	long=four-dimensional,
}

\DeclareAcronym{dD}{
	short=$d$-D,
	long=$d$-dimensional,
}

\DeclareAcronym{mD}{
	short=$m$-D,
	long=$m$-dimensional,
}

\DeclareAcronym{CT}{
	short=CT, 
	long=computed tomography,
}

\DeclareAcronym{MDCT}{
	short=MDCT, 
	long=multi-detector computed tomography,
}

\DeclareAcronym{CBCT}{
	short=CBCT, 
	long=cone-beam computed tomography,
}

\DeclareAcronym{4DCT}{
	short=4DCT, 
	long= four-dimensional computed tomography,
}

\DeclareAcronym{PET}{
	short=PET, 
	long=positron emission tomography,
}

\DeclareAcronym{SPECT}{
	short=SPECT, 
	long=single-photon emission CT,
}

\DeclareAcronym{MRI}{
	short=MRI, 
	long=magnetic resonance imaging,
}

\DeclareAcronym{MR}{
	short=MR, 
	long=magnetic resonance,
}

\DeclareAcronym{PCCT}{
	short=PCCT, 
	long=photon-counting computed tomography,
}

\DeclareAcronym{DECT}{
	short=DECT, 
	long=dual-energy computed tomography,
}

\DeclareAcronym{MBIR}{
	short=MBIR, 
	long=model-based iterative reconstruction,
}

\DeclareAcronym{WLS}{
	short=WLS,
    long=weighted least-squares,
}
\DeclareAcronym{PWLS}{
	short=PWLS, 
	long=penalized weighted least squares,
}

\DeclareAcronym{PLS}{
	short=PLS, 
	long=parallel level set,
}

\DeclareAcronym{ML}{
	short=ML, 
	long=maximum likelihood,
}

\DeclareAcronym{PML}{
	short=PML, 
	long=penalized maximum likelihood,
}

\DeclareAcronym{MLAA}{
	short=MLAA, 
	long=maximum likelihood activity and attenuation,
}

\DeclareAcronym{SPS}{
	short=SPS, 
	long=separable paraboloidal surrogates,
}

\DeclareAcronym{CS}{
	short=CS,
	long=compressive sensing,
}

\DeclareAcronym{TV}{
	short=TV,
	long=total variation,
}

\DeclareAcronym{TNV}{
	short=TNV, 
	long=total nuclear variation,
}

\DeclareAcronym{FDK}{
	short=FDK, 
	long=Feldkamp-Davis-Kress,
}

\DeclareAcronym{JTV}{
	short=JTV, 
	long=joint total variation,
}

\DeclareAcronym{DTV}{
	short=DTV, 
	long=directional total variation,
}

\DeclareAcronym{SQS}{
	short=SQS, 
	long=separable quadratic surrogate,
}

\DeclareAcronym{ADMM}{
	short=ADMM, 
	long=alternating direction method of multipliers,
}

\DeclareAcronym{DL}{
	short=DL,
	long= deep learning,
}

\DeclareAcronym{PnP}{
	short=PnP,
	long=plug-and-play,
}

\DeclareAcronym{PDF}{
	short=PDF,
	long=probability distribution function,
}

\DeclareAcronym{PSNR}{
	short=PSNR, 
	long=peak signal-to-noise ratio,
}

\DeclareAcronym{SSIM}{
	short=SSIM, 
	long=structural similarity index measure,
}

\DeclareAcronym{SNR}{
	short=SNR, 
	long=signal-to-noise ratio,
}

\DeclareAcronym{CNN}{
	short=CNN, 
	long=convolutional neural network,
}

\DeclareAcronym{NN}{
	short=NN, 
	long=neural network,
}

\DeclareAcronym{NAF}{
	short=NAF, 
	long=neural attenuation field,
}

\DeclareAcronym{NRF}{
	short=NRF, 
	long= neural radiance field,
}

\DeclareAcronym{GAN}{
	short=GAN, 
	long=generative adversarial network,
}

\DeclareAcronym{WGAN}{
	short=W-GAN, 
	long=Wasserstein GAN,
}

\DeclareAcronym{VAE}{
	short=VAE, 
	long=variational autoencoder,
}

\DeclareAcronym{MVAE}{
	short=MVAE, 
	long=multi-branch VAE,
}

\DeclareAcronym{beta-VAE}{
	short=$\beta$-VAE, 
	long=beta-variational autoencoder,
}

\DeclareAcronym{LOR}{
	short=LOR,
	long=line of response,
	long-plural-form = lines of response,
}

\DeclareAcronym{TOF}{
	short=TOF,
	long=time-of-flight,
}

\DeclareAcronym{MAP}{
	short=MAP,
	long=maximum \emph{a~posteriori},
}

\DeclareAcronym{EM}{
	short=EM,
	long=expectation-maximization,
}

\DeclareAcronym{MLEM}{
	short=MLEM,
	long=maximum-likelihood expectation-maximization,
}

\DeclareAcronym{MLE}{
	short=MLE,
	long=maximum-likelihood estimation,
}

\DeclareAcronym{OSEM}{
	short=OSEM, 
	long=ordered subsets expectation maximization,
}

\DeclareAcronym{MAPEM}{
	short=MAPEM, 
	long=maximum \textit{a posteriori} expectation maximization,
}

\DeclareAcronym{OT}{
	short=OT, 
	long=optimization transfer,
}

\DeclareAcronym{FBP}{
	short=FBP, 
	long=filtered backprojection,
}

\DeclareAcronym{IFFT}{
	short=IFFT, 
	long=inverse fast Fourier transform,
}

\DeclareAcronym{GT}{
	short=GT,
	long=ground truth,
}

\DeclareAcronym{HU}{
	short=HU,
	long=Hounsfield units,
}

\DeclareAcronym{LAC}{
	short=LAC,
	long=linear attenuation coefficient,
}

\DeclareAcronym{AC}{
	short=AC,
	long=attenuation coefficient,
}

\DeclareAcronym{MNIST}{
	short=MNIST,
	long=Modified National Institute of Standards and Technology,
}

\DeclareAcronym{LBFGS}{
	short=L-BFGS,
	long=limited-memory  Broyden-Fletcher-Goldfarb-Shanno,
}

\DeclareAcronym{KL}{
	short=KL,
	long=Kullback-Leibler
}

\DeclareAcronym{ReLU}{
	short=RelU,
	long=rectified linear unit
}

\DeclareAcronym{PSO}{
	short=PSO,
	long=particle swarm optimization
}

\DeclareAcronym{DM}{
	short=DM,
	long=diffusion model
}
\DeclareAcronym{ADM}{
	short=ADM,
	long=adaptive diffusion model
}
\DeclareAcronym{PADM}{
	short=PADM,
	long=proximal-enhanced adaptive diffusion model
}

\DeclareAcronym{LDM}{
	short=LDM,
	long=latent diffusion model
}

\DeclareAcronym{WDM}{
	short=WDM,
	long=wavelet diffusion model
}

\DeclareAcronym{DPS}{
	short=DPS,
	long=diffusion posterior sampling
}

\DeclareAcronym{DDS}{
	short=DDS,
	long=decomposed diffusion sampling
}
\DeclareAcronym{PCA}{
	short=PCA,
	long=principal component analysis 
}
\DeclareAcronym{MSE}{
	short=MSE,
	long=mean squared error 
}
\DeclareAcronym{MAE}{
	short=MAE,
	long=mean absolute error
}

\DeclareAcronym{XCAT}{
	short=XCAT,
	long=extended cardiac-torso,
}

\DeclareAcronym{OOD}{
	short=OOD,
	long=out-of-distribution,
}

\DeclareAcronym{FWHM}{
	short=FWHM,
	long=full width at half maximum,
}

\DeclareAcronym{PVE}{
	short=PVE,
	long=partial volume effect,
}

\DeclareAcronym{SDE}{
	short=SDE,
	long=stochastic differential equation,
}

\DeclareAcronym{DDPM}{
	short=DDPM,
	long=denoising diffusion probabilistic model,
}

\DeclareAcronym{DDIM}{
	short=DDIM,
	long=denoising diffusion implicit model,
}

\DeclareAcronym{MPGD}{
	short=MPGD, 
	long=manifold preserving guided diffusion,
}

\DeclareAcronym{JRAA}{
	short=JRAA,
	long=joint reconstruction of the activity and the attenuation,
}

\DeclareAcronym{DVF}{
	short=DVF,
	long=deformation vector field,
}

\DeclareAcronym{MC}{
	short=MC,
	long=motion-corrected,
}

\DeclareAcronym{DWT}{
	short=DWT,
	long=discrete wavelet transform,
}

\DeclareAcronym{iDWT}{
	short=iDWT,
	long=inverse DWT,
}

\DeclareAcronym{MCDPS}{
	short=MC-BDPS,
	long=motion-corrected reconstruction using blind DPS,
}

\DeclareAcronym{JRM}{
	short=JRM,
	long=joint reconstruction and motion estimation,
}

\DeclareAcronym{MCIR}{
	short=MCIR,
	long=moction-corrected iterative reconstruction,
}

\title{Adaptive Diffusion Models for Sparse-View Motion-Corrected Head Cone-beam CT} 
\begin{document}

\author{
    Antoine De Paepe, Alexandre Bousse, Cl\'ementine Phung-Ngoc, Youness Mellak, Dimitris Visvikis
	\thanks{
		This work used previously acquired, de-identified human CT data and did not involve any prospective data collection or interventions. The author(s) confirm(s) that all human subject research procedures and protocols are exempt from review board approval.
		}
    \thanks{
        This work was supported by CPER 2021–2027 IMAGIIS (INNOV-XS).
        }
    \thanks{
		All authors declare that they have no known conflicts of interest in terms of competing financial interests or personal relationships that could have an influence or are relevant to the work reported in this paper.
		}
    \thanks{
		All authors  are affiliated with the LaTIM, Inserm, UMR 1101, \emph{Universit\'e de Bretagne Occidentale}, Brest, France.
		} 
	\thanks{
		Corresponding author: A. Bousse, \texttt{bousse@univ-brest.fr}.
		}
	}

\maketitle

\begin{abstract}

\Ac{CBCT} is an imaging modality widely used in head and neck diagnostics due to its accessibility and lower radiation dose. However, its relatively long acquisition times make it susceptible to patient motion, especially under sparse-view settings used to reduce dose, which can result in severe image artifacts. In this work, we propose a novel framework combining \ac{JRM} with an \ac{ADM} that simultaneously addresses motion compensation and sparse-view reconstruction in head \ac{CBCT}. Leveraging recent advances in diffusion-based generative models, our method integrates a wavelet-domain diffusion prior into an iterative reconstruction pipeline to guide the solution toward anatomically plausible volumes while estimating rigid motion parameters in a blind fashion. We evaluate our method on simulated motion-affected \ac{CBCT} data derived from real clinical \ac{CT} volumes. Experimental results demonstrate that \ac{JRM}-\ac{ADM} achieves consistent quantitative improvements over both traditional and learning-based baselines. In highly undersampled cases, \ac{JRM}-\ac{ADM} improves \ac{PSNR} by more than 4~dB and \ac{SSIM} by 0.10 compared to the baseline  \ac{MC} reconstruction method. These results highlight the potential of our approach to enable motion-robust, low-dose \ac{CBCT} imaging, paving the way for improved clinical viability. The project page is available at \url{https://antoinedepaepe.github.io/jrm-adm-io/}.
	
\end{abstract}


\acresetall


\section{Introduction}

\IEEEPARstart{C}{one}-beam computed tomography (CBCT) has become a valuable imaging modality for head and neck applications. \Acs{CBCT} systems use a flat-panel detector and a cone-shaped X-ray beam to acquire volumetric images in one rotation, and are widely used in dental, sinus, and cranial imaging. In these contexts, \Ac{CBCT} often serves as a lower-dose and more accessible alternative to conventional \ac{MDCT} \cite{li_patient_2013, nicholson_novel_2021}. \Ac{CBCT} scanners are also more compact and can be deployed at the point of care (e.g., in outpatient clinics or operating rooms), offering advantages in portability and cost for head imaging \cite{rumboldt_review_2009, wu_cone-beam_2020}. While these features make \ac{CBCT} particularly appealing, continued efforts are underway to further optimize patient safety, especially in terms of radiation dose.

One strategy to further reduce radiation exposure in head \ac{CBCT} is to minimize the number of X-ray projections acquired. However, doing so inevitably introduces a trade-off between minimizing radiation dose and maintaining diagnostically sufficient image quality \cite{yan2012comprehensive}. In clinical contexts, it is essential that reconstructed images preserve critical anatomical details so that the resulting images remain clinically interpretable. Therefore, dose-reduction strategies must be carefully balanced with the need to retain key diagnostic structures.

\begin{figure}[ht]
	\centering
	  \centering
	  \includegraphics[width=\linewidth]{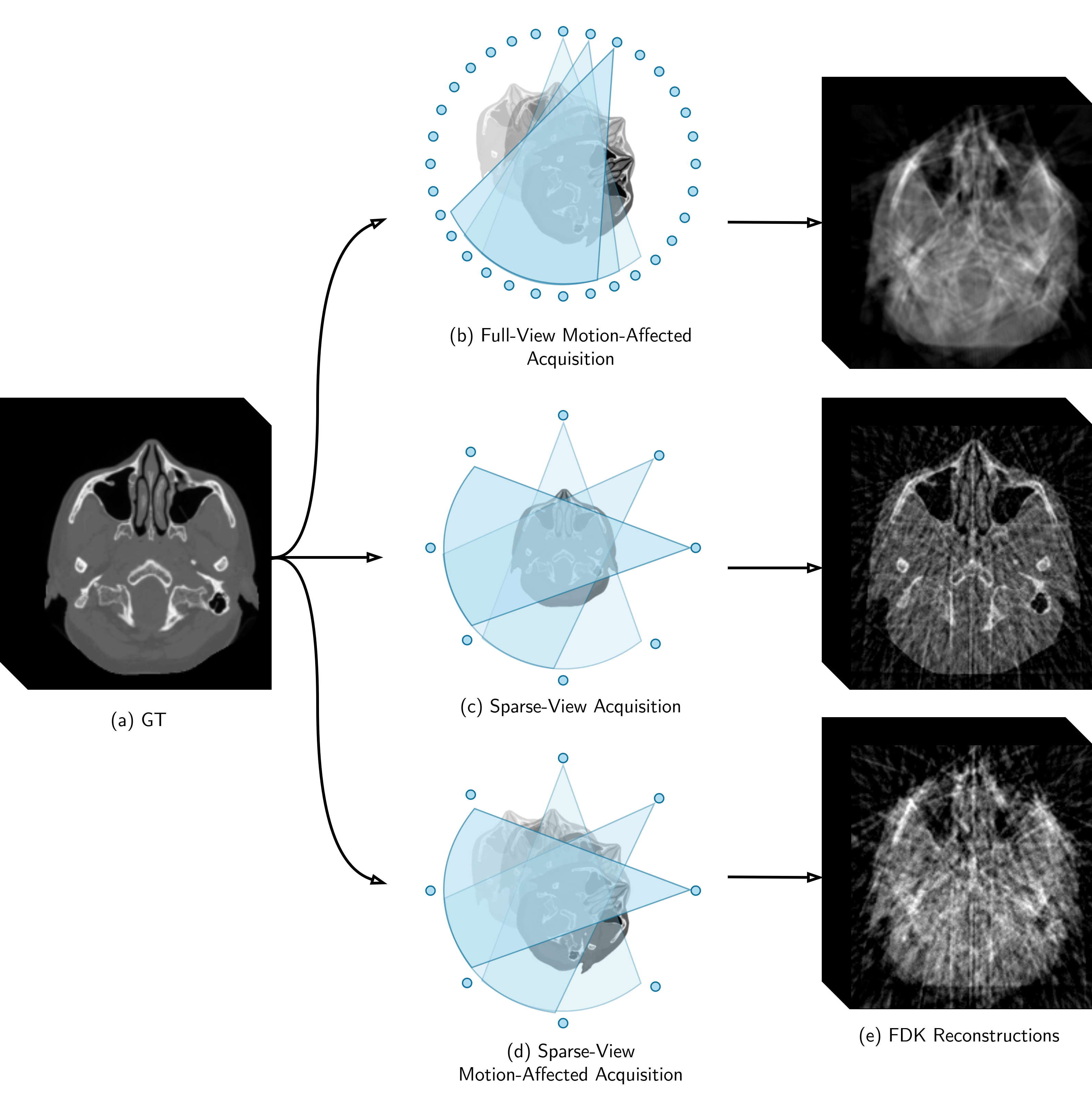}
	  \caption{Reconstructions under sparse-view setting, patient motion setting, and a combination of both.}
	  \label{fig:diff_acq}
\end{figure}

In the sparse-view context, standard algorithms—such as the \ac{FDK} \cite{feldkamp_practical_1984} algorithm—yield pronounced artifacts, leading to severe degradation of image quality, including loss of structural sharpness and distortion of clinically relevant features such as bone boundaries and soft-tissue interfaces, as shown in Figure~\ref{fig:diff_acq}. Significant research has addressed the ill-posed sparse-view reconstruction problem in \ac{CT} and \ac{CBCT}. Traditional \ac{MBIR} algorithms leverage regularization, prior information, or \ac{CS} techniques to compensate for missing projections \cite{sidky2012convex}. However, these techniques can produce over-smoothed areas that obscure subtle anatomical features and reduce the overall fidelity of the reconstructed image. More recently, \ac{DL} methods have pushed reconstruction quality further. For example, \acp{CNN} can be trained to refine \ac{FDK}-reconstructed sparse-view \ac{CBCT} images \cite{jin_deep_2017}. However, these kinds of supervised methods have drawbacks, especially when dealing with \ac{OOD} data. This limitation has motivated the use of less supervised and more robust alternatives.

On one hand, \acp{NAF} and related approaches \cite{zha_naf_2022, shin_sparse-view_2025} have been introduced for sparse-view \ac{CBCT} reconstruction. These methods leverage \acp{NRF}---originally designed for synthesizing novel views in computer graphics---to represent the \ac{CBCT} volume as an implicit neural function. On the other hand, methods at the intersection of \ac{MBIR} algorithms and \ac{DL} approaches have been developed. \Ac{PnP} algorithms address image reconstruction inverse problems by integrating a restoration \ac{NN} into iterative processes, often with theoretical convergence guarantees. While early \ac{PnP} methods used Gaussian denoisers as restoration \acp{NN}, which are not well-suited for sparse-view \ac{CBCT}, \citeauthor{leonardis_plug-and-play_2025}~\cite{leonardis_plug-and-play_2025} addressed this by training the restoration \ac{NN} to approximate a proximal operator along a defined optimization trajectory. Finally, \acp{DM} \cite{ho_denoising_2020} have emerged as state-of-the-art generative approaches capable of learning complex image distributions, and have arisen as an alternative to \acp{GAN} \cite{dhariwal2021diffusion} in medical imaging \cite{kazerouni2023diffusion}. Score-based \acp{DM} are generative models that involve training a prior \ac{PDF} by progressively converting a set of images into white noise using a \ac{SDE}. The reverse \ac{SDE} is then learned through score matching with a \ac{CNN}, allowing for image generation starting from random white noise. Building on the foundational insights of \acp{DM}, \ac{DPS} methods \cite{chung_diffusion_2022} have demonstrated remarkable capabilities in denoising, artifact suppression, and even in solving highly ill-posed reconstruction problems in low-dose and \ac{3D} sparse-view \ac{CT} \cite{chung_decomposed_2023, song_diffusionblend_2024}. Nonetheless, all these methods assume a static object and do not account for patient motion during the scan, which is often the case in \ac{CBCT} given the relatively long acquisition time \cite{venkatesh_cone_2017, wagner_quantification_2003}. In practice, patients may involuntarily move their heads, leading to substantial image artifacts, such as double contours, ghosting of edges, and an overall loss of sharpness and contrast in the reconstructed image \cite{wagner_quantification_2003}, as illustrated in Figure \ref{fig:diff_acq}.

Several strategies have been explored to obtain the motion information needed for \ac{MC} reconstruction. External tracking uses additional hardware to monitor patient motion, but requires precise registration with the imaging system. Misalignment or calibration errors can reduce correction accuracy \cite{kim2013feasibility, spin2017accuracy}. 

Alternatively, data-driven \ac{MC} reconstruction techniques can estimate the motion directly from the raw data. For example, \ac{JRM} techniques can jointly estimate the motion from the tomographic projections and reconstruct a motion-free image such that the temporal projections of the registered volume align with the measured ones \cite{bousse2015maximum, sun_iterative_2016,ouadah_self-calibration_2016, li_motion_2022}. Autofocus motion correction algorithms, which measure image quality on the reconstructed volume itself, have gained interest in recent years \cite{sisniega_motion_2017}, especially with data-driven quality metrics parameterized by \acp{NN} \cite{huang_reference-free_2022, preuhs_appearance_2020, thies_gradient-based_2025} and \acp{DM} \cite{thies2024differentiable,chen2025portable}. However, these approaches are not designed to handle sparse-view data. Moreover, in the specific case of learned quality metrics, it can be problematic when motion amplitudes differ from those seen during training.

Despite advances in sparse-view reconstruction and in motion correction, there is a conspicuous gap in methods that tackle both simultaneously, especially in the case of head \ac{CBCT}. Most prior works compartmentalize the problems: they either assume no motion when dealing with sparse-view data, or assume full-view data when correcting motion. Several  studies have explored the integration of sparse-view acquisition with motion compensation in a reconstruction scheme \cite{pengpen_motion-compensated_2015}, or through \ac{DL} approaches \cite{wu_joint_2023}. However, challenges persist---either in maintaining image quality in the former case or in extending these methods to \ac{3D} applications in the latter. To the best of our knowledge, no existing approach integrates a modern learned image prior into a \ac{JRM} framework for the specific case of head \ac{CBCT}. This motivates this work: to offer a new solution that marries the latest advances in generative image modeling with a joint motion-image optimization scheme, to fill this crucial gap.

\begin{figure*}[htbp]
	\centering
	\centering
	\includegraphics[width=\linewidth]{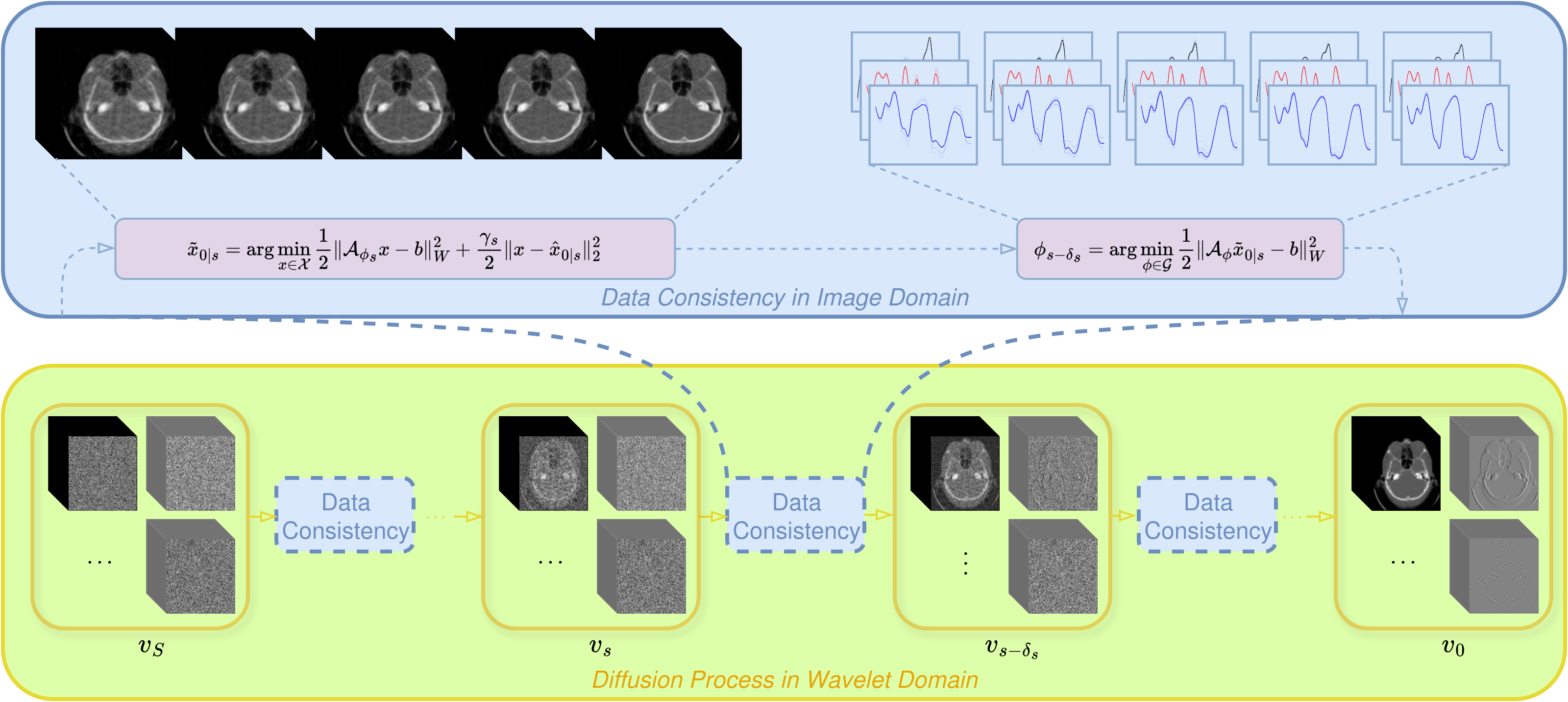}
	\caption{Overview of the proposed \acs{JRM}-\acs{ADM} approach.}
	\label{fig:adm_jrm_fig}
\end{figure*}

In this paper, we extend our previous  framework, developed for the sparse-view \ac{4D} \ac{CT} context \cite{paepe_solving_2025}, to \ac{MC} sparse-view head \ac{CBCT}, leveraging an \acf{ADM} for \ac{DPS}. Our approach, namely \ac{JRM}-\ac{ADM}, integrates a \ac{DPS} scheme inspired by \citeauthor{chung_decomposed_2023}~\cite{chung_decomposed_2023} into an iterative reconstruction algorithm, where the \ac{DM} acts as a probabilistic regularizer to guide the solution toward anatomically plausible images and the motion is blindly estimated (Figure~\ref{fig:adm_jrm_fig}). This joint estimation process allows our algorithm to iteratively refine both the image and the motion parameters until convergence is achieved, thereby reducing motion-induced artifacts while preserving fine anatomical details. To reduce the memory footprint, we employ a \ac{WDM} which learns a \ac{PDF} defined on a compressed space such that the training of the \ac{DM} is more memory-efficient.  

The remainder of this paper is organized as follows: Section~\ref{sec:method} details the theoretical framework and integration of \ac{JRM} and \ac{JRM}-\ac{ADM}; Section~\ref{sec:results}  describes the experimental setup and evaluation metrics, and presents the results; Section~\ref{sec:discussion} discusses clinical implications and future research directions; finally, Section~\ref{sec:conclusion} concludes this work.

\section{Materials and Methods}\label{sec:method}

In the following, all vectors are represented in columns. `$\transp$' denotes the matrix transpose operation. For a given column vector $\boldz = [z_1,z_2,\dots,z_n]\transp \in \R^n$, $[\boldz]_i$ denotes the $i$th entry of $\boldz$, i.e., $[\boldz]_i \triangleq z_i$. $[\boldA;\boldB]$ denotes the vertical concatenation of two matrices $\boldA$ and $\boldB$ with equal number of columns. Given a positive-definite square matrix $\boldA$, we define the norm $\|\cdot\|_{\boldA}$ by $\|\boldz\|_{\boldA} \triangleq \boldz\transp \boldA \boldz$ where $\boldz$ is a vector of same dimensions as the number of columns in $\boldA$. $\boldzero_{\calZ}$ and $\boldI_{\calZ}$ are respectively the zero element and the identity operator in the real vector space $\calZ$. The \ac{3D} attenuation image to reconstruct is represented by a vector $\boldx \in \calX\triangleq \R^m $ with $m = n_{x} \cdot n_{y} \cdot {n_z}$ voxels.

\subsection{Problem Formulation: Joint  Reconstruction and Motion Compensation}\label{sec:jrm}

A \ac{CBCT} scan operates by rotating an X-ray source and a detector around the patient's head, capturing a collection of \ac{2D} projections from various angles $\theta_1,\theta_2,\dots,\theta_{n_\rma}$, with ${n_\rma}$ being the number of angles, and we denote by $\boldcalR_k \colon \calX \to \calY$ the $k$th X-ray line integral operator (i.e., corresponding to angle $\theta_k$), where $\calY \triangleq \R^n$ denotes the projection space with $n=n_\rmh \cdot n_\rmw$ such that each vector of $\calY$ represents an $n_\rmh \times n_\rmw$ projection image. The $k$th X-ray transmission measurement data is a random vector $\boldy_k = [y_{1,k},\dots,y_{n,k}]\transp\in\calY$, which, in the absence of electronic noise, follows a Poisson distribution with independent entries given by
\begin{equation}\label{eq:poisson_nomotion} 
	(y_{i,k} \mid \boldx) \sim \mathrm{Poisson}(\ybar_{i,k}(\boldx))
\end{equation}
where $\ybar_{i,k}(\boldx) \triangleq \mathbb{E}[y_{i,k}|\boldx]$ is the conditional expected number of detected photons given $\boldx$, which is given by the Beer--Lambert law (assuming the X-ray source is monochromatic):
\begin{equation}\label{eq:bll_nomotion} 
	\ybar_{i,k}(\boldx) = I \cdot \rme^{- [\boldcalR_k (\boldx)]_i  }  \, .
\end{equation}
Additionally,  the $y_{i,k}$s are conditionally independent given $\boldx$.  Given the entire measurement  $\boldy \triangleq [\boldy_1;\dots;\boldy_{n_{\rma}}] \in \calY^{n_{\rma}}$, image reconstruction can be achieved by finding the \ac{MAP} estimate of $\boldx$, i.e.,
\begin{equation}\label{eq:map_nomotion}
	\max_{\boldx \in \calX} \, p(\boldy \mid  \boldx) \cdot p(\boldx)
\end{equation} 
where the conditional \ac{PDF}  $p(\boldy |  \boldx)$ is given by \eqref{eq:poisson_nomotion} and $p(\boldx)$ is a prior \ac{PDF} on $\boldx$, which is generally unknown and replaced (in its post-log form) by a regularizer promoting piecewise-smooth images. Solving \eqref{eq:map_nomotion} can be achieved with an \ac{OT} algorithm \cite{elbakri2002statistical} when $p(\boldx)$ is differentiable, or with primal-dual algorithms \cite{sidky2012convex}  when $p(\boldx)$ is not differentiable (e.g., \ac{TV}).

The acquisition of the data $\boldy \triangleq [\boldy_1;\dots;\boldy_{n_{\rma}}] \in \calY^{n_{\rma}}$ is performed over a time interval $[0,T]$ during which the head is subject to motion, thus creating inconsistencies  between the projections $\boldy_k$. In the following, we assume there is no intra-projection motion (i.e., each $\boldy_k$ is an instant snapshot of the patient's head), and we denote by $t_k$, $k=1,\dots,n_{\rma}$, the time at which the $k$th projection occurs, with $t_1=0$ and $t_{n_{\rma}} = T$. At each time $t\in[0,T]$, the support of the attenuation image $\boldx$ is affected by a \ac{3D} rigid motion given by a time-dependent parameter $\boldomega(t) = [ \omega_1(t),\dots,\omega_6(t) ]$  
\begin{align}
	[\omega_1(t),\omega_2(t),\omega_3(t)] = {} & [\vartheta_x(t),\vartheta_y(t),\vartheta_z(t)] \triangleq \boldvartheta(t) \nonumber \\
	[\omega_4(t),\omega_5(t),\omega_6(t)] = {} & [\tau_x(t),\tau_y(t),\tau_z(t)] \triangleq \boldtau(t) \nonumber
\end{align}
where $\boldvartheta(t)\in[0,2\pi]^3$ is the vector of the rotation angles around each axis and $\boldtau(t)\in\R^3$ is the translation vector. Furthermore, $\boldcalW_{\boldomega(t)} \colon \calX \to \calX$ denotes the image-to-image transformation (i.e., a warping matrix) associated with the rigid transformation of parameter $\boldomega(t)$.

To preserve motion smoothness, we introduce a temporal regularization by modeling each component of $\boldomega(t)$  using cubic B-splines: 
\begin{equation}\label{eq:bspline}
	\omega_{p}(t) = \sum_{i=1}^{n_\rmc} \phi_{i, p} B\left(\frac{t-s_i}{r}\right) \quad \forall p=1,\dots,6
\end{equation}
where $B \colon \R \longrightarrow \R$ is the cubic B-spline basis function, $s_i$, $i=1,\dots,n_\rmc$ are the $n_\rmc$ uniformly distributed control points with $s_1=0$ and $s_{n_\rmc}=T$ (not to be confused with the acquisition times $t_k$, $k=1,\dots,n_\rma$), $r$ is the distance between knots and $\boldphi = \{\phi_{i,p}\}_{i,p=1}^{n_\rmc,6}$ regroups all the motion  parameters. Furthermore, we define $\boldcalW_{\boldphi}^k \colon \calX \to \calX$ as the image-to-image transformation operator of parameter $\boldphi$ at time $t_k$ corresponding to the $k$th \ac{CBCT} projection, i.e.,
\begin{equation}
	\boldcalW_{\boldphi}^k \triangleq \boldcalW_{\boldomega(t_k)} \nonumber
\end{equation}
where $\boldomega$ is defined as \eqref{eq:bspline}.

With the introduction of the motion defined by $\boldphi$, the forward model given by \eqref{eq:poisson_nomotion} and \eqref{eq:bll_nomotion} can be redefined as
\begin{equation}\label{eq:poisson_motion} 
	(y_{i,k} \mid \boldx,\boldphi) \sim \mathrm{Poisson}(\ybar_{i,k}(\boldx,\boldphi))
\end{equation}
with $\ybar_{i,k}(\boldx,\boldphi) \triangleq \mathbb{E}[y_{i,k}|\boldx,\boldphi]$ defined as
\begin{equation}\label{eq:bll_motion} 
	\ybar_{i,k}(\boldx,\boldphi) = I \cdot \rme^{- \left[\boldcalR_k \circ \boldcalW_{\boldphi}^k  (\boldx)\right]_i  } \, .
\end{equation}
In the absence of a prior on $\boldphi$, \ac{MC} reconstruction of the image $\boldx$ from the measurement $\boldy$ can be achieved by performing \ac{JRM} through a \ac{MAP} optimization problem
\begin{equation}\label{eq:map_motion}
	\max_{\boldx \in \calX, \boldphi \in \calC}  \, p(\boldy \mid \boldx,\boldphi) \cdot p(\boldx)
\end{equation}
where the conditional \ac{PDF} $p(\boldy | \boldx, \boldphi)$ is given by \eqref{eq:poisson_motion} and \eqref{eq:bll_motion}, and $\calC = \R^{n_\rmc \times 6}$ denotes the set of control points $\boldphi$ modeling the rigid motion. Defining the motion-incorporated projector $\boldcalA_{\boldphi}$ as 
\begin{equation}
	\boldcalA_{\boldphi} \triangleq \left[\boldcalR_1 \circ \boldcalW_{\boldphi}^1;\dots;\boldcalR_{n_\rma} \circ \boldcalW_{\boldphi}^{n_\rma} \right] \colon \calX \to \calY^{n_\rma} \, , \nonumber
\end{equation}
a \ac{WLS} approximation of the negative log-likelihood  $-\log p(\boldy | \boldx, \boldphi)$ is \cite{elbakri2002statistical}
\begin{equation}\label{eq:wls}
	-\log p(\boldy \mid \boldx, \boldphi) \approx \frac{1}{2} \left\| \boldcalA_{\boldphi}(\boldx) - \boldb\right\|^2_{\boldW} + C
\end{equation}
where $C$ is independent of $\boldx$ and $\boldphi$, $\boldb \triangleq [\boldb_1;\dots;\boldb_{n_\rma}]$, $\boldb_k \triangleq [b_{1,k},\dots,b_{n,k}]\transp$ with $b_{i,k} \triangleq \log (I/y_{i,k})$ (assuming $y_{i,k}>0$ for all $i,k$) and $\boldW = \mathrm{diag}[\boldy]$, and \eqref{eq:map_motion} is approximated in its post-log form as 
\begin{equation}\label{eq:pwls_jrm}
	\min_{\boldx \in \calX, \boldphi \in \calC}  \, \left\| \boldcalA_{\boldphi}(\boldx) - \boldb\right\|^2_{\boldW} + \beta R(\boldx)
\end{equation}
where $R\colon\calX\to\R$ is a convex regularizer playing the role of $-\log p$ and $\beta>0$ controls the strength of $R$. Solving \eqref{eq:pwls_jrm} is a standard \ac{JRM} optimization problem which can be found in the literature in different forms \cite{bousse2015maximum, sun_iterative_2016,ouadah_self-calibration_2016, li_motion_2022}.

\subsection{Joint Reconstruction and Motion Estimation with Diffusion Models}

In the following, $s=1,\dots,S$ denotes the diffusion index. 

\subsubsection{Background on Diffusion Models}

In the absence of a tractable prior \ac{PDF} $p(\boldx)$, $\boldx$ can be sampled via a model trained through diffusion. 

A commonly adopted approach is the  \ac{DDPM} \cite{ho_denoising_2020}, which samples $\boldx_s$ given $\boldx_{s-1}$, $s=1,\dots,S$, starting from an initial image $\boldx_0$ sampled from the training dataset with \ac{PDF} $p^\mathrm{data}$, 
\begin{equation}\label{eq:xtfromxtm1}
	\boldx_s = \sqrt{\alpha_s} \boldx_{s-1} +   \sqrt{1-\alpha_s}  \boldepsilon_s
\end{equation}
where $\boldepsilon_s \sim \scrN \left( \boldzero_\calX , \boldI_\calX  \right) $ and $\alpha_s$ is a scaling factor decreasing from $\alpha_0 = 1$ to $\alpha_S = 0$  such that $\boldx_S \sim \calN (\boldzero_{\calX},\boldI_{\calX})$. One prominent sampling algorithm, namely \ac{DDIM} \cite{song2020denoising}, approximates the reverse process to sample an image from a generalized version of $p^\mathrm{data}(\boldx)$ that approximates the theoretical prior $p(\boldx)$, with the following update rule:
\begin{align}\label{eq:ddimsampling}
	\boldx_{s - 1} = {} &   \sqrt{\bar{\alpha}_{s - 1}} \boldxhat_{0|s} + \sqrt{1 - \bar{\alpha}_{s - 1} - \sigma_{s}^2} \cdot \frac{\boldx_{s} -  \sqrt{\bar{\alpha}_{s}} \boldxhat_{0|s}}{\sqrt{1 - \bar{\alpha}_{s}}}  \nonumber \\
	& + \sigma_s \boldepsilon_{s} \, , \quad \boldepsilon_{s} \sim  \scrN(\boldzero_{\calX},\boldI_{\calX})
\end{align}
where the specific case of $\sigma_s = \sqrt{\left(1-\bar{\alpha}_{s-1}\right)/\left(1-\bar{\alpha}_{s}\right)}$ $\cdot \sqrt{1-\bar{\alpha}_{s}/\bar{\alpha}_{s-1}}$ leads to \ac{DDPM} sampling, $\bar{\alpha}_s = \prod_{r=1}^s \alpha_r$  and $\boldxhat_{0|s} \triangleq \mathbb{E}[\boldx_0|\boldx_s]$ is given by Tweedie's formula,
\begin{equation}
	\mathbb{E}[\boldx_0\mid \boldx_s]=  \frac{1}{\sqrt{\bar{\alpha}_s}} (\boldx_s + (1 - \bar{\alpha}_s) \nabla \log p_s(\boldx_s) ) \, , \nonumber
\end{equation}
$p_s$ being the \ac{PDF} of $\boldx_s$. In practice, we use \ac{DDIM} with a step $\delta_s > 1$, so \eqref{eq:ddimsampling} is implemented as an update from $s$ to $s-\delta_s$. The score function $\nabla_{\boldx_s}\log p_s(\boldx_s)  $ is intractable and therefore $\boldxhat_{0|s}$ is approximated through a \ac{NN} $\hat{\boldx}_{\boldeta}  \colon \calX \times [0,S]\to \calX$ with parameter vector $\boldeta$ trained to recover $\boldx_{0}$ from $\boldx_{s}$ as 
\begin{equation}\label{eq:x0_matching}
	\min_{\boldeta} \, \mathbb{E}_{s, \boldx_0, \boldx_{s}} \left[ \left\| \hat{\boldx}_{\boldeta}(\boldx_s,s) - \boldx_{0} \right\|_2^2 \right],
\end{equation}
where $s \sim \scrU[0, S]$, $\boldx_0 \sim p^\mathrm{data}$, and $\boldx_{s} \sim \scrN(\sqrt{\bar{\alpha}_s}\boldx_0, (1-\bar{\alpha}_s)\boldI_{\calX})$.


\subsubsection{Diffusion Posterior Sampling for Blind Inverse Problems}


\Acp{DM} can be used for sparse-view \ac{CBCT} \ac{MC} image reconstruction via \ac{DPS}, with incorporation of a rigid motion parametrized by $\boldphi_s$ (cf. \eqref{eq:bspline}) and by replacing the conditional expectation $\mathbb{E}\left[ \boldx_{0} |\boldx_s\right]$ with $\mathbb{E}\left[ \boldx_{0} |\boldx_s, \boldphi_s, \boldy\right]$. This can be achieved with the conditional score which can be derived from Bayes' formula, assuming $\boldphi_s$ and $[\boldx_0,\dots,\boldx_S]$ are independent (which is an acceptable assumption for rigid motion):
\begin{align}
	\nabla_{\boldx_s} \log p(\boldx_s \mid \boldy,\boldphi_s)  = {} & \nabla_{\boldx_s} \log p_s(\boldx_s) \nonumber \\
	& + \nabla_{\boldx_s} \log p(\boldy \mid \boldx_s,\boldphi_s) \, \nonumber
\end{align} 
where $\boldy$ is dependent on $\boldx_s$ through the sequence $\boldx_{s-1}, \boldx_{s-2}, \dots, \boldx_0 = \boldx$, as specified by the \ac{DM} in \eqref{eq:xtfromxtm1} and the forward model described in \eqref{eq:poisson_motion} and \eqref{eq:bll_motion}, with $\boldphi = \boldphi_s$. 

The \ac{DPS} framework can use the following approximation   \cite{chung_diffusion_2022}:
\begin{align}\label{eq:dps}
	\nabla_{\boldx_s} \log p(\boldy \mid \boldx_s,\boldphi_s) \approx  \nabla_{\boldx_s} \log p(\boldy \mid \boldxhat_{0|s}(\boldx_{s}),\boldphi_s)
\end{align}	
where  the log-conditional \ac{PDF} $ \log p(\boldy | \boldxhat_{0|s},\boldphi_s) $ is given by  \eqref{eq:wls} with $\boldx = \boldxhat_{0|s}$.  However, computing such a gradient is not always possible, especially in large-scale inverse problems due to the memory footprint. Alternative solutions have been proposed, such as the manifold-preserving guided diffusion shortcut approach \cite{he2023manifold}, which consists in replacing $\nabla_{\boldx_s} \log p(\boldy | \boldxhat_{0|s}(\boldx_{s}) , \boldphi_s)$ with $\nabla_{\boldxhat_{0|s}} \log p(\boldy | \boldxhat_{0|s},\boldphi_s)$.
This shortcut does not require backpropagating through the \ac{NN} $\boldxhat_\eta$ used to approximate $\boldxhat_{0|s}$. A similar approximation was proposed in \citeauthor{jiang2024strategies}~\cite{jiang2024strategies}.

Another approximation can be derived from the conditional version of Tweedie's formula
\begin{equation}\label{eq:tweedie_cond}
	\mathbb{E}\left[ \boldx_{0} |\boldx_s, \boldphi_s, \boldy\right] = 
     \mathbb{E}\left[ \boldx_{0} | \boldx_s\right] \nonumber\\ 
	+ \frac{1-\bar{\alpha}_s}{\sqrt{\bar{\alpha}_s}}\nabla_{\boldx_s} \log p(\boldy | \boldx_s,\boldphi_s) \, . \nonumber
\end{equation}
Combined with \eqref{eq:dps} and the manifold-preserving guidance, this formula resembles a gradient ascent step starting from $\boldxhat_{0|s} = \mathbb{E}\left[ \boldx_{0} | \boldx_s\right]$ to maximize $\log p(\boldy | \boldx , \boldphi)$. This observation motivates the utilization of the following proximal approximation \cite{zhu_denoising_2023,chung_decomposed_2023}:
\begin{align}
	\mathbb{E}\left[ \boldx_{0} \mid \boldx_s, \boldphi_s, \boldy\right] \approx {} & \argmin_{\boldx\in \calX} \frac{1}{2}\| \boldcalA_{\boldphi_s}(\boldx) - \boldb\|^{2}_{\boldW}  \nonumber \\ 
	{} & + \frac{\gamma_s}{2} \| \boldx - \hat{\boldx}_{0|s} \|^2_{2}  \nonumber \\
	\triangleq {}  & \tilde{\boldx}_{0|s}  \, .\label{eq:prox}
\end{align}
where the parameter $\gamma_s>0$ dictates the strength of the diffusion prior, which has to be carefully chosen.

Meanwhile, the motion parameter $\boldphi_{s-1}$ can be updated by solving
\begin{equation}\label{eq:motion_update}
	 \boldphi_{s-1} = \argmin_{\boldphi\in\calC} \, \frac{1}{2} \| \boldcalA_{\boldphi} (\tilde{\boldx}_{0|s}) - \boldb  \|_{\boldW}^2\, 
\end{equation}
which can be solved with any iterative algorithm initialized from  $\boldphi_{s}$. Similarly to the image update, a step size $\delta_s>1$ can be used.

\subsubsection{Diffusion Models in Wavelet Transform Domain}

Applying \acp{DM} to \ac{3D} medical imaging poses significant challenges due to high computational cost and  memory footprint. To mitigate these issues, latent \acp{DM} operate in a compressed space. Recent studies have proposed performing the diffusion process in the wavelet domain \cite{friedrich2024wdm}, which substantially reduces memory requirements during both training and inference while achieving state-of-the-art performance.

In this context, an orthogonal \ac{DWT} $\boldcalV \colon \calX  \to \calZ \triangleq \mathbb{R}^{8 \times \frac{n_x}{2} \times \frac{n_y}{2} \times \frac{n_z}{2}}$, with  $\boldcalV^{-1} = \boldcalV\transp$, is employed to decompose the \ac{3D} image $\boldx$ into an eight-channel wavelet coefficient image $\boldv$ that has half the spatial resolution in each dimension compared to $\boldx$. This representation allows processing the eight smaller channels in parallel, thus reducing the computational time and the memory cost.  

The orthogonality of $\boldcalV$ ensures a lossless transition between the wavelet domain $\calZ $ and the original image domain $\calX$. Specifically, the diffusion process described in \eqref{eq:xtfromxtm1} can be reformulated in the wavelet domain as
\begin{equation}\label{eq:v_diff}
	\underbrace{\boldcalV (\boldx_{s})}_{\boldv_{s}} = \sqrt{\alpha_s} \, \underbrace{\boldcalV(\boldx_{s-1})}_{\boldv_{s-1}} + \sqrt{1-\alpha_s} \, \underbrace{\boldcalV (\boldepsilon_s)}_{\tilde{\boldepsilon}_s}, \nonumber
\end{equation}
with $\boldepsilon_s \sim \scrN(\boldzero_{\calX}, \boldI_{\calX})$ so that $\tilde{\boldepsilon}_s \triangleq \boldcalV \boldepsilon_s \sim \scrN(\boldzero_{\calZ}, \boldI_{\calZ})$ as $\boldcalV$ is orthogonal.

Thus, training a \ac{NN} $\hat{\boldv}_{\boldeta} \colon \calZ \times [0,S] \to \calZ$ can be achieved in a similar fashion as in \eqref{eq:x0_matching} (in the wavelet domain $\calZ$)
to predict $\boldv_0$ from $\boldv_s$ and can be utilized to predict $\boldx_0$ from $\boldx_s$ through $\boldcalV^{-1}$.

\subsubsection{Algorithm Summary}

The overall approach, namely \ac{JRM}-\ac{ADM}, is summarized in Algorithm~\ref{algo:blindmgpd}. Similarly to \citeauthor{song2020denoising}~\cite{song2020denoising}, we used a \ac{DDIM} approach with a time step $\delta_s>1$ to speed up the sampling. The subproblems in \eqref{eq:prox} and \eqref{eq:motion_update} (corresponding to steps \ref{eq:prox1} and \ref{eq:prox2} in Algorithm~\ref{algo:blindmgpd}) are solved using iterative algorithms initialized from $\hat{\boldx}_{0|s}$ and $\boldphi_s$ respectively.

Following the estimation of $\tilde{\boldx}_{0|s}$ by the proximal step, the wavelet coefficient vector $\tilde{\boldv}_{0|s}$ is extracted from $\tilde{\boldx}_{0|s} $ as
\begin{equation}
	\tilde{\boldv}_{0|s} = \boldcalV(\tilde{\boldx}_{0|s})
\end{equation}
and the new wavelet coefficient vector $\boldv_s$ is sampled using the \ac{DDIM} update \eqref{eq:ddimsampling} with  $\tilde{\boldv}_{0|s}$ replacing $\boldxhat_{0|s}$ and $\sigma_s = 0$ as proposed in \citeauthor{song2020denoising}~\cite{song2020denoising}.

Additionally, we implemented an accelerated version using the jumpstart strategy, namely \ac{JRM}-\ac{ADM}\textsubscript{js}, proposed in \citeauthor{jiang2024strategies}~\cite{jiang2024strategies}, using a pre-estimated image and motion by solving \eqref{eq:pwls_jrm} without regularizer. The wavelet coefficients were then extracted from the pre-estimated image and degraded following the diffusion process \eqref{eq:xtfromxtm1} (in the wavelet space) for $S'$ steps with $S'<S$, then used as an initialization for $\boldv_s$ instead of white noise.

\begin{algorithm}
	\caption{Pseudo code for \ac{JRM}-\ac{ADM}.}\label{algo:blindmgpd}
	\begin{algorithmic}[1]
		\Require $S$, $\boldy$, $\{ \gamma_s \}_{s=0}^{S}$, $\{ \delta_s \}_{s=0}^{S}$, $\{ \alpha_s \}_{s=0}^{S}$
		\State $s \gets S$
        \State $\boldepsilon \sim \calN(\boldzero_{\calZ}, \boldI_{\calZ})$
        \State $\boldv_{s} \gets \boldepsilon$
		\State $\boldphi_{s} \gets \boldzero_{\calC}$

        \While{$s > 0$}
            \State $\hat{\boldv}_{0|s} \gets \hat{\boldv}_{\boldeta}(\boldv_s, s)$
            \State $\hat{\boldx}_{0|s} \gets \boldcalV^{-1}(\hat{\boldv}_{0|s})$
            \State $\tilde{\boldx}_{0|s} \leftarrow \argmin_{\boldx\in \calX} \frac{1}{2}\| \boldcalA_{\boldphi_{s}}(\boldx) - \boldb\|^{2}_{\boldW} + \frac{\gamma_s}{2} \| \boldx - \hat{\boldx}_{0|s} \|^2_{2} $  \label{eq:prox1}
            \State $\boldphi_{s - \delta_s}  \leftarrow  \argmin_{\boldphi\in\calC} \, \frac{1}{2} \| \boldcalA_{\boldphi} (\tilde{\boldx}_{0|s}) - \boldb  \|_{\boldW}^2$ \label{eq:prox2}
            \State $\tilde{\boldv}_{0|s} \gets \boldcalV(\tilde{\boldx}_{0|s})$
            \State $\boldv_{s - \delta_s} \gets \sqrt{\bar{\alpha}_{s - \delta_s}} \tilde{\boldv}_{0|s} + \sqrt{1 - \bar{\alpha}_{s- \delta_s}}\cdot\frac{\boldv_{s} -  \sqrt{\bar{\alpha}_{s}} \tilde{\boldv}_{0|s}}{\sqrt{1 - \bar{\alpha}_{s}}}$
            \State $s = s - \delta_s$
		\EndWhile
        \State $\boldx_0 = \boldcalV^{-1}(\boldv_0)$
		\State \textbf{return} $\boldx_0,\boldphi_{0}$
	\end{algorithmic}
\end{algorithm}
\section{Results}\label{sec:results}

\newlength{\tempdima}
\setlength{\tempdima}{0.36\linewidth}

\newlength{\tempdimah}
\setlength{\tempdimah}{0.36\linewidth}

\newlength{\tempwidth}
\setlength{\tempwidth}{2.0\columnwidth}

\newcommand{\textcolormetrics}{lightgray}

\subsection{Data Preparation}

\input{./figures/bigfig1}

We utilized the CQ500 dataset \cite{chilamkurthy2018deep}, which comprises 491 \ac{MDCT} scans collected from patients, some with symptoms of head trauma or stroke. From the full dataset, we selected a subset of 296 volumes satisfying two main criteria: (1) a sufficient number of axial slices, and (2) a small slice thickness, ensuring appropriate resolution for simulating \ac{CBCT} acquisitions. The selected volumes are clipped to the \ac{HU} range $[-1000, 2000]$ and divided into training, validation, and test sets with 263, 15, and 18 samples, respectively. The validation set is used during model training and to fine-tune sampling hyperparameters.

We trained the \ac{NN} $\hat{\boldv}_{\boldeta}$ using the \textsc{Adam} optimizer for approximately 1.2 million iterations, using rotations and translations for data augmentation. The training was performed on volumes normalized to the range $[-1, 1]$.  This normalization was explicitly incorporated into the forward model \eqref{eq:bll_motion} while performing \ac{JRM}-\ac{ADM}.

We simulated realistic motion-affected \ac{CBCT} acquisitions following \eqref{eq:poisson_motion} and \eqref{eq:bll_motion} with \ac{GT} volumes $\boldx = \boldx^\star$ from the testing dataset and random \ac{GT} motion B-spline parameters $\boldphi= \boldphi^\star$ tuned to generate a continuous motion following \eqref{eq:bspline} (with $n_{\rmc} = 20$ control points) with translations in the range of $[-5, 5]$~mm  and rotations in the range of $[-5, 5]$ degrees (the same motion was used for all images in the testing dataset) and $I = 5 \cdot 10^5$.

Each simulation consisted of 120 projections uniformly distributed across  $[0,2\pi]$ using the cone-beam projector implemented in the Carterbox GitHub repository \footnote{\url{https://github.com/carterbox/torch-radon}} (a fork of TorchRadon \cite{ronchetti2020torchradon}). The acquisition geometry included a source-to-isocenter distance of 785~mm, a source-to-detector distance of 1,200~mm, with 500\texttimes{}700 detector pixels and an isotropic pixel size of 0.5~mm. From these projections, we uniformly sampled the desired number of projection angles $n_\rma\in\{20,60\}$ for reconstruction, allowing comparison of the same motion under different sparse-view settings. Reconstructions were performed on a \ac{3D} grid with isotropic voxel spacing of 1\texttimes{}1\texttimes{}1~mm, within volumes of size 160\texttimes{}192\texttimes{}192 voxels. 

\subsection{Experimental settings}

\input{./figures/bigfig2}

All parameters were finely tuned to optimize the metrics.

For performance comparison, we used standard \ac{FDK} reconstruction as a baseline. 

As a model-based approach, we employed \ac{JRM} \eqref{eq:pwls_jrm} using \ac{TV} regularization for $R$. The regularization strength $\beta$ is adjusted according to the number of projections $n_\rma$, with $\beta = 7.5 \cdot 10^2$ and $3 \cdot 10^3$ for $n_\rma = 60$ and $20$, respectively.

For the full \ac{JRM}-\ac{ADM} pipeline, we set $S = 1000$ diffusion steps and $\delta_s = 10$. During the first 10 steps, we disabled the diffusion regularization by setting $\gamma_s = 0$ across all sampling conditions. In this initialization phase, the image and motion estimation steps (i.e., steps \ref{eq:prox1} and \ref{eq:prox2} in Algorithm~\ref{algo:blindmgpd}) are performed with 15 \textsc{RMSprop} steps. For the remaining diffusion steps $s$, we set $\gamma_s = 9 \cdot 10^3$ and $1.1 \cdot 10^4$ for $n_\rma = 60$ and $n_\rma = 20$, respectively, and we used five \textsc{RMSprop} iterations for the image and motion updates.

We then implemented \ac{JRM}-\ac{ADM}\textsubscript{js} with $S' = 400$ and $\delta_s = 10$. The initial estimation of the volume and motion parameters is performed by solving \eqref{eq:pwls_jrm} using an alternating optimization scheme over the volume $\boldx$ and the motion parameters $\boldphi$. Each sub-problem is optimized using \textsc{RMSprop} with 15 gradient steps, repeated for 10 alternating iterations. The diffusion regularization strength $\gamma_s$ was fixed over time and set according to the number of projections, with $\gamma_s = 2 \cdot 10^4$ and $6 \cdot 10^4$ for $n_\rma = 60$ and $20$, respectively.

To ensure temporal alignment across different sparse-view scenarios, we selected the central projection angle, indexed by $k_0$, as the reference frame, corresponding to the acquisition angle $ \theta_{k_0}=\pi$. We then compared  the warped reconstructed images $\boldcalW_{\boldphi}^{k_0} \boldx$ (except for \ac{FDK} which has no motion compensation) with the \ac{GT} image $\boldcalW_{\boldphi^\star}^{k_0} \boldx^\star$ using \ac{PSNR} and \ac{SSIM}. We also report on the accuracy of both translations and rotations to evaluate the accuracy of the estimated motion, computed as the \ac{MAE} between the estimated motion $\omega_p(t_k)$, $p=1,\dots,6$ and $k=1,\dots,n_\rma$, and the true motion used to generate the data.

\subsection{Preliminary Results}

We first conducted a preliminary experiment to assess the effect of motion in a \ac{DPS} framework. For this experiment, we considered a 60-view motion-affected acquisition and performed reconstruction with a standard \ac{DPS} approach, i.e., Algorithm~\ref{algo:blindmgpd} with $\boldphi = \boldzero_\calC$, for different values of $\gamma_s = \gamma$ (which we chose to be constant). 

\begin{figure}
	\centering
	\includegraphics[width=\columnwidth]{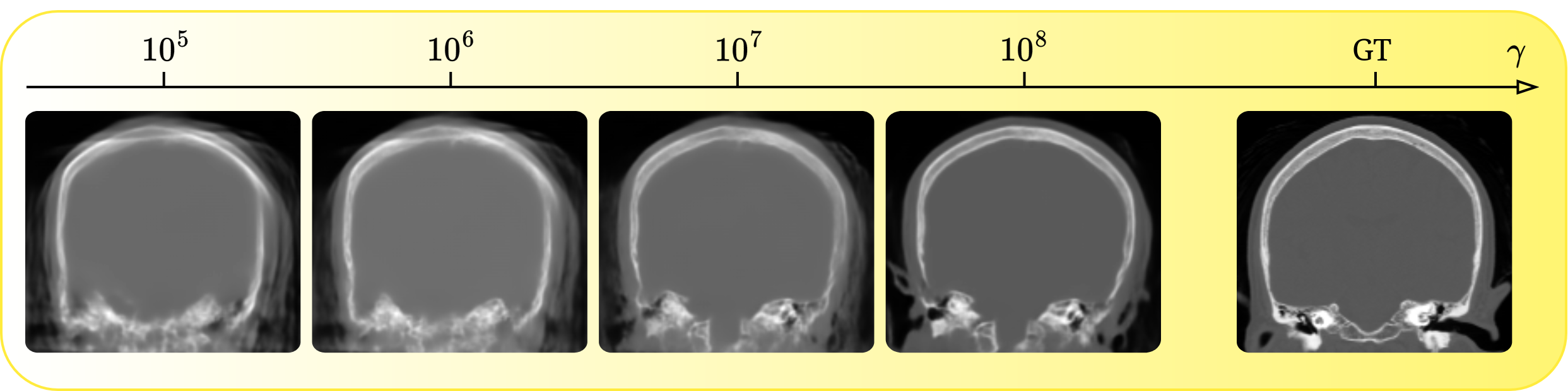}
	\caption{\ac{GT} and reconstructed volumes with  non-\ac{MC} \ac{DPS} approach for different values of $\gamma$, in the 60-view setting. When $\gamma$ is low, the algorithm tends to ignore the prior and instead focuses on the  data which are affected by motion, thus resulting in motion artifacts. On the contrary, when $\gamma$ is high, the algorithm tends to favor the prior, which produces a motion-free image that deviates from the \ac{GT}.} 
	\label{fig:no_mc_dps_gt}
\end{figure}

The results are shown in Figure~\ref{fig:no_mc_dps_gt}. When $\gamma$ is low, the algorithm largely ignores the prior and relies heavily on the motion-affected data, leading to motion artifacts in the reconstructed image, including blurring, double contours, and distortions of the overall structure. Conversely, when $\gamma$ is high, the algorithm places more weight on the prior, resulting in a motion-free image that deviates from the \ac{GT} as the reconstruction tends to ignore the data.

These observations highlight the importance of explicitly modeling and compensating for motion as part of the \ac{DPS} reconstruction pipeline to achieve accurate results.

\subsection{Sparse-view Experimental Results}

\subsubsection{Fracture-free Head Images}

Figure~\ref{fig:recon1} and Figure~\ref{fig:recon2} present the reconstructed images and the \ac{GT} for $n_\rma=60$ and $n_\rma=20$ angles of view, respectively, on two different patients with no apparent head trauma.


The \ac{FDK}-based reconstructions exhibit noise and pronounced streak artifacts, especially for the 20-view acquisition, where the reduction in number of views leads to a significant loss of structural information. Additionally, motion artifacts degrade image quality, leading to a reduction in fine details.

The \ac{JRM}-\ac{TV} reconstructions are capable of compensating for motion and addressing streak artifacts in the 60-view setting, resulting in sharp and high-quality images. However, in the 20-view setting, the reconstructions exhibit  both increased streak artifacts and over-smoothed areas, leading to a general loss of detail, with fine structures becoming less discernible. 

The \ac{JRM}-\ac{ADM}\textsubscript{js} and \ac{JRM}-\ac{ADM} methods exhibit similarly  high levels of detail in both the 60-view and 20-view scenarios, significantly outperforming \ac{JRM}-\ac{TV} in terms of accuracy, as reflected by higher \ac{PSNR} and \ac{SSIM} values as shown in Table~\ref{tab1}. Fine details in the nasal region  are accurately preserved and motion is fully compensated.

However, in the more challenging 20-view case, \ac{JRM}-\ac{ADM} seems to outperform \ac{JRM}-\ac{ADM}\textsubscript{js}, with the latter showing a tendency to hallucinate structures. While motion is generally well estimated---supported by low \ac{MAE} in both translations and rotations---this degradation can be attributed to an insufficient number of refinement steps in \ac{JRM}-\ac{ADM}\textsubscript{js}, which becomes particularly limiting under the highly ill-posed conditions of 20-view motion-affected reconstructions.

In addition, closer examination of the reconstructed volumes (Figure~\ref{fig:recon1}, last row) reveals subtle structural inconsistencies emerging in specific brain regions. These artifacts likely stem from over-smoothing effects induced by the diffusion-based \ac{NN}, especially in low-contrast areas where the network tends to suppress fine anatomical variations. This behavior suggests that the diffusion prior may compromise local structural fidelity under such conditions.

\subsubsection{Fractured Head Image}

\input{figures/fractured_head}

In this experiment, we used the attenuation image of a patient with several fractures (cf. magnified areas in Figure~\ref{fig:broken_bone}).

In both 60- and 20-view settings, the \ac{FDK} method performs poorly, preventing reliable diagnosis. 


Under the 60-view setting, all \ac{JRM}-based reconstructions appear accurate and all fractures are visible.  



In the more challenging 20-view case, the upper fracture is visible  for all \ac{JRM}-based reconstructions (cf. blue-magnified area) but the lower fracture  (cf. yellow-magnified area)  is only visible with \ac{JRM}-\ac{ADM}.


\subsubsection{Quantitative Analysis}

Figure~\ref{fig:ssim_psnr_60} and Figure~\ref{fig:ssim_psnr_20} show the \ac{PSNR} and \ac{SSIM} for the 60-view and 20-view settings, respectively, for the entire testing dataset. While in the 60-view setting the distributions appear fairly similar (with the exception of two samples for which \ac{JRM}-\ac{TV} is outperformed by \ac{JRM}-\ac{ADM} and \ac{JRM}-\ac{ADM}\textsubscript{js}), a difference is visible between the three methods in the 20-view setting, with \ac{JRM}-\ac{TV} clearly underperforming and \ac{JRM}-\ac{ADM} performing the best.

\begin{figure}[htbp]
	\centering
	\begin{tikzpicture}
		\begin{axis}[
			scatter/classes={
				JRMTV={mark=square*,blue},
				JRMADMJS={mark=triangle*,green!60!black},
				JRMADM={mark=diamond*,orange}
			},
			xlabel={SSIM},
			ylabel={PSNR},
			ylabel style={yshift=-10pt},
			legend style={font=\small, at={(0.2,0.99)}, anchor=north, legend columns=1}, 
			width=0.9\columnwidth,
			height=0.6\columnwidth,
			grid=major,
			scatter,
			only marks,
			label style={font=\small}, 
			tick label style={font=\small} 
			]
			
\addplot[scatter, scatter src=explicit symbolic]
coordinates {
    (0.9759, 34.75) [JRMTV]
    (0.8976, 26.43) [JRMTV]
    (0.9555, 31.51) [JRMTV]
    (0.7720, 22.19) [JRMTV]
    (0.9670, 34.07) [JRMTV]
    (0.9296, 28.95) [JRMTV]
    (0.9657, 32.93) [JRMTV]
    (0.9179, 26.62) [JRMTV]
    (0.9744, 33.20) [JRMTV]
    (0.8111, 21.73) [JRMTV]
    (0.9322, 27.80) [JRMTV]
    (0.9470, 28.70) [JRMTV]
    (0.9573, 32.55) [JRMTV]
    (0.9136, 25.63) [JRMTV]
    (0.8995, 26.97) [JRMTV]
    (0.9624, 32.62) [JRMTV]
    (0.9784, 35.25) [JRMTV]
    (0.9620, 30.33) [JRMTV]
};

\addplot[scatter, scatter src=explicit symbolic]
coordinates {
    (0.9761, 35.67) [JRMADMJS]
    (0.9722, 33.88) [JRMADMJS]
    (0.9716, 34.83) [JRMADMJS]
    (0.9266, 28.33) [JRMADMJS]
    (0.9771, 36.28) [JRMADMJS]
    (0.9814, 36.44) [JRMADMJS]
    (0.9504, 31.12) [JRMADMJS]
    (0.9558, 30.75) [JRMADMJS]
    (0.9747, 35.97) [JRMADMJS]
    (0.9271, 26.92) [JRMADMJS]
    (0.8759, 24.57) [JRMADMJS]
    (0.9704, 32.95) [JRMADMJS]
    (0.9388, 29.40) [JRMADMJS]
    (0.9772, 34.10) [JRMADMJS]
    (0.8605, 24.79) [JRMADMJS]
    (0.9754, 35.55) [JRMADMJS]
    (0.9569, 33.30) [JRMADMJS]
    (0.9667, 31.48) [JRMADMJS]
};

\addplot[scatter, scatter src=explicit symbolic]
coordinates {
    (0.9813, 37.38) [JRMADM]
    (0.9689, 32.81) [JRMADM]
    (0.9564, 31.56) [JRMADM]
    (0.9766, 34.46) [JRMADM]
    (0.9780, 36.22) [JRMADM]
    (0.9439, 30.16) [JRMADM]
    (0.9506, 30.73) [JRMADM]
    (0.9418, 28.42) [JRMADM]
    (0.9799, 37.73) [JRMADM]
    (0.9399, 27.88) [JRMADM]
    (0.8855, 25.05) [JRMADM]
    (0.9172, 25.83) [JRMADM]
    (0.8652, 23.83) [JRMADM]
    (0.9712, 32.02) [JRMADM]
    (0.8568, 24.57) [JRMADM]
    (0.9640, 32.31) [JRMADM]
    (0.9707, 35.34) [JRMADM]
    (0.9156, 26.40) [JRMADM]
};

			\legend{JRM-TV, JRM-ADM\textsubscript{js}, JRM-ADM }
		\end{axis}
	\end{tikzpicture}
	\caption{\Ac{SSIM} vs \ac{PSNR} scatter plot for the 60-view reconstructions using JRM-TV, JRM-ADM\textsubscript{js} and JRM-ADM, each dot  corresponding to a reconstructed image of the 18 volumes from the testing dataset.}
	\label{fig:ssim_psnr_60}
\end{figure}
\begin{figure}[htbp]
	\centering
	\begin{tikzpicture}
		\begin{axis}[
			scatter/classes={
				JRMTV={mark=square*,blue},
				JRMADMJS={mark=triangle*,green!60!black},
				JRMADM={mark=diamond*,orange}
			},
			xlabel={SSIM},
			ylabel={PSNR},
			ylabel style={yshift=-10pt},
			legend style={font=\small, at={(0.2,0.99)}, anchor=north, legend columns=1}, 
			width=0.9\columnwidth,
			height=0.6\columnwidth,
			grid=major,
			scatter,
			only marks,
			label style={font=\small}, 
			tick label style={font=\small} 
			]
			
\addplot[scatter, scatter src=explicit symbolic]
coordinates {
    (0.8982, 28.18) [JRMTV]
    (0.7520, 21.83) [JRMTV]
    (0.6996, 21.52) [JRMTV]
    (0.5808, 17.65) [JRMTV]
    (0.7514, 22.94) [JRMTV]
    (0.7891, 23.48) [JRMTV]
    (0.9052, 28.57) [JRMTV]
    (0.7885, 21.76) [JRMTV]
    (0.9220, 28.45) [JRMTV]
    (0.6809, 17.91) [JRMTV]
    (0.9131, 28.81) [JRMTV]
    (0.8302, 22.99) [JRMTV]
    (0.8908, 27.73) [JRMTV]
    (0.7895, 20.89) [JRMTV]
    (0.8924, 28.69) [JRMTV]
    (0.8702, 26.32) [JRMTV]
    (0.9037, 28.37) [JRMTV]
    (0.8506, 24.33) [JRMTV]
};

\addplot[scatter, scatter src=explicit symbolic]
coordinates {
    (0.9365, 31.15) [JRMADMJS]
    (0.8720, 26.07) [JRMADMJS]
    (0.9053, 28.87) [JRMADMJS]
    (0.7861, 23.05) [JRMADMJS]
    (0.9020, 28.68) [JRMADMJS]
    (0.9078, 28.58) [JRMADMJS]
    (0.8915, 28.35) [JRMADMJS]
    (0.8657, 25.09) [JRMADMJS]
    (0.9336, 30.38) [JRMADMJS]
    (0.8221, 22.57) [JRMADMJS]
    (0.8758, 26.33) [JRMADMJS]
    (0.8977, 26.21) [JRMADMJS]
    (0.8849, 27.36) [JRMADMJS]
    (0.8927, 25.63) [JRMADMJS]
    (0.8265, 24.79) [JRMADMJS]
    (0.9174, 29.57) [JRMADMJS]
    (0.8780, 27.68) [JRMADMJS]
    (0.9059, 27.27) [JRMADMJS]
};

\addplot[scatter, scatter src=explicit symbolic]
coordinates {
    (0.9557, 33.17) [JRMADM]
    (0.9405, 30.46) [JRMADM]
    (0.9148, 28.95) [JRMADM]
    (0.9444, 30.91) [JRMADM]
    (0.9383, 31.63) [JRMADM]
    (0.9050, 28.26) [JRMADM]
    (0.9001, 28.10) [JRMADM]
    (0.9076, 26.94) [JRMADM]
    (0.9658, 35.04) [JRMADM]
    (0.9277, 27.44) [JRMADM]
    (0.8792, 25.87) [JRMADM]
    (0.9213, 26.94) [JRMADM]
    (0.8033, 22.47) [JRMADM]
    (0.9396, 28.86) [JRMADM]
    (0.8608, 25.97) [JRMADM]
    (0.9503, 32.25) [JRMADM]
    (0.9294, 30.24) [JRMADM]
    (0.8991, 26.39) [JRMADM]
};

			\legend{JRM-TV, JRM-ADM\textsubscript{js}, JRM-ADM }
		\end{axis}
	\end{tikzpicture}
	\caption{\Ac{SSIM} vs \ac{PSNR} scatter plot for the 20-view  reconstructions using JRM-TV, JRM-ADM\textsubscript{js} and JRM-ADM, each dot  corresponding to a reconstructed image of the 18 volumes from the testing dataset.}
	\label{fig:ssim_psnr_20}
\end{figure}

Figure~\ref{fig:motion_patterns_60} and Figure~\ref{fig:motion_patterns_20} show the estimated motion $\omega_p(t_k)$, $p=1,\dots,6$ and $k=1,\dots,n_\rma$, for the 60-view and 20-view settings, respectively, for two different patients. While motion estimation seems accurate for all methods for 60-view data, we observe that \ac{JRM}-\ac{ADM} and \ac{JRM}-\ac{ADM}\textsubscript{js} tend to yield more accurate motion estimates compared to \ac{JRM}-\ac{TV} in the 20-view setting. This improved estimation facilitates the diffusion process in refining fine image details, contributing to overall reconstruction quality.

\input{./figures/motion1}
\input{./figures/motion2}

Finally, Table~\ref{tab1} summarizes all the metrics averaged over all images in the testing dataset.

\begin{table}[H]
	\scriptsize
	\centering
	\begin{tabular}{lcccccc}
		\toprule
		$n_\rma$ & Method & \ac{PSNR} $(\uparrow)$ & \ac{SSIM} $(\uparrow)$ & \ac{MAE} $\tau$ $(\downarrow)$ & \ac{MAE} $\vartheta$ $(\downarrow)$ \\
		\midrule
		\multirow{4}{*}{60} 
		& FDK & 18.36 & 0.34 & -    & -    \\
		& JRM-TV & 29.57 & 0.93 & 0.37 & 0.11 \\
		& JRM-ADM\textsubscript{js} & \underline{\textbf{32.02}} & \underline{\textbf{0.95}} & \underline{\textbf{0.18}} & 0.12 \\
		& JRM-ADM & 30.71 & 0.94 & 0.29 & \underline{\textbf{0.10}} \\
		\midrule
		\multirow{4}{*}{20} 
		& FDK & 16.57 & 0.22 & -    & -    \\
		& JRM-TV & 24.47 & 0.82 & 0.94 & 0.18 \\
		& JRM-ADM\textsubscript{js} & 27.09 & 0.88 & 0.36 & 0.19 \\
		& JRM-ADM & \underline{\textbf{28.88}} & \underline{\textbf{0.92}} & \underline{\textbf{0.32}} & \underline{\textbf{0.13}} \\
		\bottomrule
	\end{tabular}
		\caption{Quantitative results of methods in comparison on the CQ500 dataset for the \ac{MC} reconstruction tasks for $n_\rma = 60$ and $n_\rma = 20$. The best values are highlighted in bold.}
	\label{tab1}
\end{table}

\subsection{Reconstruction Time and Resource Utilization}

In terms of reconstruction performance, Table \ref{tab:computation} clearly shows that \ac{JRM}-\ac{TV} is the slowest method in both the 20-view and 60-view cases. This is due to its requirement of a much larger number of alternating solving iterations to reach convergence, which significantly increases the total reconstruction time.

\begin{table}[h]
    \scriptsize
    \centering
    \begin{tabular}{lccc}
        \toprule
        $n_\rma$ & Method & Reconstruction Time (s) & GPU Footprint (GB)\\
        \midrule
        \multirow{4}{*}{60} 
        & FDK & 0.04 & 0.02 \\
        & JRM-TV & 1261.72 & 15.8 \\
        & JRM-ADM\textsubscript{js} & 590.46 & 16.2  \\
        & JRM-ADM & 1026.61 & 16.2 \\
        \midrule
        \multirow{4}{*}{20} 
        & FDK & 0.03 & 0.02 \\
        & JRM-TV & 397.98 & 4.7  \\
        & JRM-ADM\textsubscript{js} & 205.02 & 5.1 \\
        & JRM-ADM & 359.42 & 5.1 \\
        \bottomrule
    \end{tabular}
    \caption{Performance comparison of reconstruction methods: time and GPU footprint.}
    \label{tab:computation}
\end{table}

Among the \ac{MC} reconstruction methods, \ac{JRM}-\ac{ADM}\textsubscript{js} is the fastest, achieving a reduction of approximately 42\% in reconstruction time compared to \ac{JRM}-\ac{ADM} in both 20-view and 60-view cases. Note that we took into account the pre-estimation time required by \ac{JRM}-\ac{ADM}\textsubscript{js}. It is also worth noting the per-iteration computational cost: solving a single iteration of \eqref{eq:prox} takes 0.114~s for 20 views  and 0.712~s for 60 views, while solving \eqref{eq:motion_update} requires 0.548~s for 20 views and 1.196~s for 60 views. These values illustrate the scaling of computational effort with the number of views.

All our experiments were performed on an NVIDIA RTX 6000 Ada Generation. Regarding GPU usage, all the \ac{MC} reconstruction methods exhibit a moderate footprint of around 5--16~GB, increasing with the number of views.

\section{Discussion}\label{sec:discussion}

This study highlights the potential of \acp{ADM} for \ac{MC} sparse-view head \ac{CBCT} imaging. That said, several limitations still need to be addressed before this approach can be fully applied in real-world clinical settings. 

While our experiments simulate \ac{CBCT} data from these \ac{MDCT} volumes, \ac{CBCT} and \ac{MDCT} differ in several ways. In particular, \ac{CBCT} is generally more susceptible to scatter and exhibits higher and more complex noise levels, but leads to lower dose compared to \ac{MDCT}. To bridge this gap and enable realistic inference in actual \ac{CBCT} scenarios, additional techniques may be required. For example, scatter estimation methods \cite{ritschl2015robust, lorenzon2025joint} could be employed to compensate for inherent scatter, while test-time adaptation approaches \cite{barbano2025steerable} could account for differences in attenuation levels between the \ac{MDCT} data used for training and real \ac{CBCT} acquisitions. 
 
Although the \ac{CT} volumes we used are based on real acquisitions, we do not have direct access to the corresponding raw projection data. Hence, while we simulated realistic conditions such as Poisson noise and sparse-view acquisitions, we did not  account for real-world effects, such as scatter or beam hardening. These factors can impact image quality and reconstruction fidelity in clinical practice, highlighting the need to validate our method on real projection data.

One major challenge lies in the computational cost of the method. For example, running a 60-view \ac{JRM}-\ac{ADM} diffusion process takes about 17 minutes, which is not suitable for urgent clinical scenarios where rapid diagnosis is crucial, particularly in life-threatening situations. Our approach is currently better suited for follow-up examinations.

To address this limitation, two potential strategies can be considered. First, motion estimation could be performed at a lower resolution than that of the reconstructed volume, which would reduce computational cost but may result in the loss of fine structural details. Second, incorporating differentiable X-ray rendering techniques—such as DiffDRR \cite{gopalakrishnan2022fast}—could improve computational efficiency by removing the need for explicit volume warping. More recently, \citeauthor{jiang2025differentiable}~\cite{jiang2025differentiable} proposed a differentiable forward and back-projector for rigid motion estimation, further advancing efficiency in motion-compensated reconstruction.

In this work, we focused exclusively on full-range HU images ($-1000$ to $2000$~HU), such as those used for imaging cranial fractures. This choice is consistent with previous work on motion-corrected head \ac{CBCT} by other groups \cite{thies_gradient-based_2025}. However, bone fractures are often accompanied by internal bleeding, which may be visible on \ac{CBCT} images acquired with different imaging settings but is also affected by motion. Further work is needed to assess the impact of motion correction on the detection and quantification of internal bleeding.

\section{Conclusion}\label{sec:conclusion}

In this work, we addressed sparse-view head \ac{CBCT} image reconstruction with motion compensation. While prior research has made substantial progress in either sparse-view reconstruction or motion correction independently, our approach is the first to combine these two aspects within a single framework in the context of head \ac{CBCT}, building upon our previous work on \ac{MC} sparse-view 4DCT. By integrating a diffusion-based prior in a blind fashion, we enable robust reconstruction across arbitrary geometries and rigid motion patterns, while preserving fine anatomical details even under severely sparse acquisition conditions. Our proposed framework opens a promising path toward safer, lower-dose, and artifact-resistant \ac{CBCT} imaging for the head, which can potentially be implemented in routine clinical practice.

\AtNextBibliography{\footnotesize}
\printbibliography
\end{document}